\documentclass{PoS}
\newcommand{\rotten}{\mbox{OH\,231.8+4.2}}
\newcommand{\rottenn}{\mbox{OH\,231.8}}
\newcommand{\sect}{\mbox{$''$}}
\newcommand{\degr}{\hbox{$^\circ$}}
\newcommand{\kms}{\mbox{km\,s$^{-1}$}}
\newcommand{\secp}{\mbox{\rlap{.}$''$}}
\newcommand{\juc}{\mbox{$J$=1$-$0}}
\newcommand{\jdu}{\mbox{$J$=2$-$1}}

\DeclareGraphicsExtensions{.png,.pdf,.jpg,.jpeg}

\title{Resolving discrepancy in the pPN \rotten}

\ShortTitle{Resolving discrepancy in the pPN \rotten}

\author{\speaker{J.-F. Desmurs}\\
        Observatorio Astron\'omico Nacional, Spain\\
        E-mail: \email{desmurs@oan.es}}
\author{J. Alcolea\\
        Observatorio Astron\'omico Nacional, Spain\\
        E-mail: \email{j.alcolea@oan.es}}
\author{M. Lindqvist\\
        Department of Space, Earth and Environment,
        Chalmers University of Technology, Onsala Space
        Observatory, 439 92 Onsala, Sweden\\
        E-mail: \email{michael.lindqvist@chalmers.se}}
\author{V. Bujarrabal\\
        Observatorio Astron\'omico Nacional, Spain\\
        E-mail: \email{v.bujarrabal@oan.es}}
\author{R. Soria-Ruiz\\
        Observatorio Astron\'omico Nacional, Spain\\
        E-mail: \email{r.soria@oan.es}}
\author{P. de Vicente \\
        Observatorio de Yebes (IGN), Yebes, Spain\\
        E-mail: \email{p.devicente@oan.es}}

\abstract{\rotten\ is an archetypal pre-planetary nebulae (pPN).  It's
  a binary system surrounded by bipolar nebula. Some years ago the
  authors extensively studied it and performed several VLBI
  observations from which they obtained mas-resolution maps of the SiO
  (7\,mm) and H$_2$O (1.3\,cm) maser emissions (see \cite{Desmurs04}).
  H$_2$O masers were found to be distributed in two areas along the
  symmetry axis of the nebulae oriented nearly north-south delineating
  a bipolar outflow and their astrometric positions were accurately
  measured.  SiO masers, indicating the position of the Mira component
  of the binary system, form a structure perpendicular to the axis of
  the nebulae.  The general picture of the source looked satisfactory,
  except for the relative position of the two masers.  Surprisingly,
  SiO masers, were tentatively placed 250~mas away (370~AU) from the
  apparent center of the outflow.  Using the ALMA we observed the SiO
  maser emission at 86\,GHz and accurately derived the position of the
  Mira component.  Combining our previous VLBA data and our new ALMA
  observations we are now able to give a more complete and detailed
  description of the inner part of this amazing pPN.}

\FullConference{14th European VLBI Network Symposium \& Users Meeting (EVN 2018)\\
		8-11 October 2018\\
		Granada, Spain}

\begin{document}

\section{Introduction} 
Planetary nebulae (PNe) evolve from the envelopes of asymptotic giant
branch (AGB) stars, through the very short ($\sim$1000\,yr) phase of
pre-planetary nebula (pPN).  During this phase, the nebular morphology
and kinematics are dramatically altered: the spherical, slowly
expanding AGB envelope becomes a PN, usually with axial symmetry and
high axial velocities.  This spectacular metamorphosis is thought to
result from the interaction of a fast and highly collimated bipolar
wind, ejected in the late-AGB or early post-AGB phase, with the slow
spherical AGB envelope.  To explain the onset of these bipolar flows
several models have postulated the presence of binary companions, an
accretion disk, rotating magnetic fields or a combination of these (see
e.g. \cite{Balick02}; \cite{Soker02}; \cite{Frank04}).
A spectacular case of a pPN is the Rotten Egg Nebula, a.k.a. \rotten, a
55\sect\ long bi-lobed nebula around the binary star QX\,Pup.
\rotten\ (hereafter \rottenn) is a well studied pPN over a wide range
of wavelengths. The distance to this source, 1.54 +0.02 -0.01 kpc
(\cite{Choi12}), and the inclination of the bipolar axis with respect
to the plane of the sky, $\sim{36\degr}$, are well known. The central
source is a binary system formed by the primary, a M9 III Mira
variable (i.e. AGB star), and an A0 main sequence companion (see
\cite{SanchezC04}). The presence of a late-type star in the core of a
bipolar post-AGB nebula like \rottenn\  is very unusual since the
central stars of pPNe are typically hotter with spectral types from A
to K. This led some authors to suggest that this could be a
born-again object\,: a star that after having initiated its post-AGB
evolution, comes back to the AGB as a result of the last thermal pulse
(see \cite{Alcolea01} and references therein). This remarkable bipolar
nebula shows all the signs of post-AGB evolution: fast bipolar outflows
with velocities $\sim$200--400~\kms, shock-excited gas and
shock-induced chemistry (see \cite{SanchezC00}; \cite{Alcolea01};
\cite{Bujarrabal02}; \cite{Bujarrabal12}; \cite{SanchezC15}).  In spite
of this, \rottenn\  also presents SiO and H$_2$O maser emission,
which can be used to probe the very inner environment of this pPN with
the highest resolution.
  
Although the central sources remains directly unseen, we know that it
is a binary system, as the scattered light shows the presence a cool
M9III Mira variable primary and a weaker A0V secondary (see
\cite{SanchezC04}). There is also evidence for various sites of
activity, as shown by the misalignment between the symmetry axis of CO
and that of other images (see \cite{GomezR01}), and also by the
0,25\secp separation tentatively found between the locations of the SiO
(see \cite{SanchezC00}) and 22 GHz H$_2$O masers (see \cite{Choi12}).
It has been suggested that the presence of yet another star could
explain the difference in size between the north and south lobes\,: if
there is another mass-losing AGB star close to the engine of \rottenn,
this would introduce an asymmetry in the ambient medium the post-AGB
outflow is running into (\cite{Riera05}).  However the exact location
of these different sources and their role in the post-AGB evolution of
the system remains unclear, mostly due to the lack of sub-arc second
resolution images of the molecular gas in the central parts of the
nebula.

About 15 years ago, we carried out several VLBA projects to study the
maser emission from this source. We observed the 7\,mm SiO maser $v$=2,
\juc\ emission (\cite{SanchezC02}) and obtained ~0.3\,mas-resolution
VLBA maps of the 3\,mm SiO maser $v$=1, \jdu\ emission (see
\cite{Desmurs04}).
However, we were not able to determine an accurate
absolute position for the SiO emission with subsequent phase
referencing VLBI observations.

Our observations revealed that the SiO emission arises from several
spots, less than $\sim$10$^{13}$\,cm in size, forming a structure
elongated in a direction perpendicular to the symmetry axis of the
nebula occupying a region of about 8\,mas (12\,AU) in extension and a
few mas across. Such a different distribution from the spherical shells
of spots usually found in evolved stars (see e.g$.$ VLBA observations
by \cite {Diamond94}; \cite{Desmurs99}), seems to reveal the presence
of an equatorial torus of gas around the primary Mira star with a
radius of $\sim$6~AU. The velocities found along the torus are
consistent with the keplerian rotation around the central star(s)
(\cite{SanchezC02}).

In addition we also mapped the H$_2$O maser emission which was found to
arise from several compact spots distributed in mainly two areas about
$30$~mas each in size (see Fig. 2, \cite{Desmurs07}) and separated by
about $60$~mas in the direction of the symmetry axis of the nebula.
Their absolute positions and the proper motions of the individual spots
were measured very accurately. The northern area corresponds to the
blue-shifted emission and the southern area to the red-shifted one. The
region in which the water masers are distributed seems to delineate the
walls of a double cone outflow perpendicular to the equatorial disk
traced by the SiO masers.

The emerging picture seems consistent for the results of the
relative position of the two maser emissions. The published results on
the SiO emission (\cite{SanchezC02}), tentatively place
the Mira primary about 250~mas away from the center of the H$_2$O
outflow.  This separation results in a lower limit distance between the
two centers of maser emission at least 370\,AU, far too much for
an interacting binary system.

During the last 20 years the SiO masers have been slowly fading away.
At the epoch of our first mm-VLBI observations SiO masers were much
brighter, with a peak flux of 20\,Jy, but today their intensity has
dropped down by a factor of nearly 40 (in fifteen years), and the peak
of the emission is only $\sim$ 0.5\,Jy. Given its
sensitivity and high resolution ALMA is the only instrument that could
provide both the sensitivity and spatial resolution needed to
disentangle the question of the absolute position of the SiO masers.

\section{Observations and Results}

We observed \rottenn\ with ALMA in Band 3 (86~GHz) as part of the Cycle
4 program (ID: 2016.1.00472.S; PI: J.-F. Desmurs).  We used the
extended configuration of ALMA (C40-9) to reach the highest spatial
resolution available ($\sim$0.\sect{1}).  The observation was conducted
using 41 12m antennas with baselines ranging from 21 to 14851\,m
between 23\&24 September, 2017.  The first run failed and the second
gave a total time on-source of 21.77 minutes. The precipitable water
vapor (PWV) was 0.78 mm during the observations.

The correlator was configured to simultaneously observed SiO $v$=1
\jdu\ and $v$=2 \jdu\ with narrow filters of 60MHz ($\sim$210~\kms, the
total width of the lines emission covers about 15~\kms) split in 960
channels to reach a high spectral resolution ($\sim$0.22~\kms) to
properly sample the individual maser components (typically 1 or
2~\kms).  In addition, we also added two basebands with a broad band
width (total aggregate bandwidth $\sim$~3.6\,GHz) to detect the
continuum and other spectral lines at lower velocity resolution.

\begin{figure}[h]
 \begin{center}
  \includegraphics[scale=0.5]{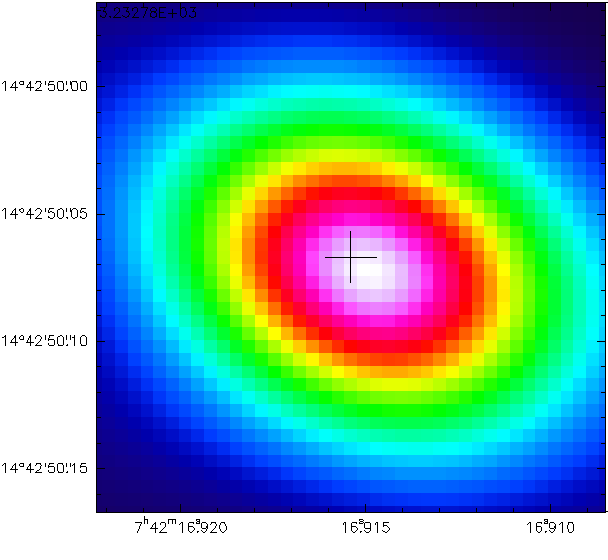}
  \caption{\em{Continuum map at 87GHz of the \rotten\ . Beam
      resolution is 80x50mas, rms=0.07\,mJy, the integrated flux
      density is 3.405$\pm$0.075 mJy. The cross mark the phase center
      position ICRS 07:42:16.9154 -14.42.50.067}}
  \label{fig1}
 \end{center}
\end{figure}

For pointing, bandpass and flux calibration we regularly observed
quasar J0750+1231 and for phase calibration J0746-1555.
The data were calibrated using the Common Astronomy Software
Applications (CASA, \cite{McMullin07}) package, version
4.7.2. Following the data reduction scripts provided by ALMA we applied
the standard bandpass, flux, and gain calibrations. Based on the phase
and amplitude variations on calibrators we estimated an absolute flux
calibration uncertainty of about 10\%. Self-calibration was performed
to improve the signal-to-noise ratio.  The continuum and SiO line
images were finally created from the calibrated visibilities using the
CASA tclean task with Briggs weighting and the parameter 'robust' set
to 0.5. Maps of 2.5x2.5arcsec were produced with a restoring beam of
0.1x0.07arcsec FWHM (corresponding to a physical scale of 150x105\,AU
at a distance of 1.54\,kpc).
The rms noise level of the image is $\sim$5\,mJy/beam per channel.
We also produced maps with uniform weighting improving our restoring
beam 0.06x0.04\,arcsec (setting the parameter ``robust=-2'' in tclean).
Absolute positions can be determined up to an accuracy of about the
synthesized HPWB divided by twice the signal-to-noise ratio (SNR);
The map of the continuum emission (with a RMS=7.17e-02\,mJy) shows a
single components slightly elongated in a direction consistent with the
axis of the nebulae (see Fig. \ref{fig1}). Modeling the visibility data
with the uvmultifit (see \cite{MartiV14}) yields a total flux density
of 3.5$\pm$0.06mJy at position RA=07:42:16.914980$\pm$0.000033\,s
DEC=-014:42:50.072631$\pm$0.000369\,arcsec.
\begin{figure}[h]
 \begin{center}
  \includegraphics[scale=0.5]{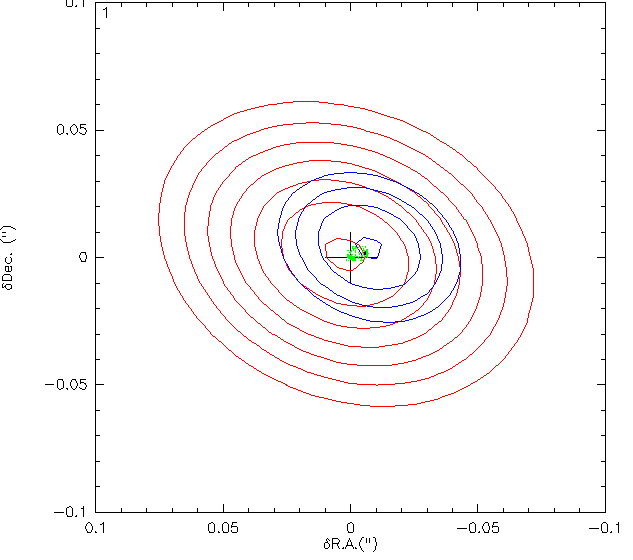}
  \caption{\em{Map of the blue-shifted (in blue) with an integrated
      flux density of 563.87$\pm$0.44 mJy and red-shifted with an
      integrated flux density of 1.46081$\pm$0.00086\,Jy (red) $v$=1,
      \jdu\ SiO emission at 3~mm detected toward \rotten. The beam
      resolution is 60x40mas. In green, the VLBA map of SiO $v$=2,
      \juc\ observed in 2002 (\cite{SanchezC02}). The relative position
      of ALMA and VLBA maps is just indicative. The cross mark the
      phase center position ICRS 07:42:16.9154 -14:42:50.067}}
  \label{fig2}
 \end{center}
\end{figure}

For SiO, apart the thermal line ($v$=0, \jdu), only the $v$=1
\jdu\ maser was detected. Figure \ref{fig2} shows the two components
detected with respectively a blue-shifted peak of about
566.4$\pm$3.8\,mJy/beam and a red-shifted peak of about 1454.9$\pm$
6.6\,mJy/beam (value extracted by fitting a Gaussian component with the
task uvmultifit) separated by a distance of about $\sim$11\,mas.
The absolute position measured for the center of emission of the
components is RA=07:42:16.91522847$\pm$7.7x10$^{-7}$s
DEC=-014:42:50.06400164$\pm$1.044x10$^{-5}$arcsec in agreement with the
position recently published by \cite{Dodson18} (using the KVN).  This
result places without any ambiguity the SiO emission in between the two
areas of water maser emission, suggesting that the binary system is
indeed a close one, and that the launching of the jets from the
interaction between the two components is indeed possible.

\section{Conclusion} 
Using ALMA, we detected and measured the SiO maser $v$=1, \jdu\ and the
thermal SiO line ($v$=0, \jdu). We also detected the continuum emission
which traces the binary system, and that is spatially coincident with
the SiO emission. We confirmed the SiO position tracing the Mira
component of the binary system as measured by \cite{Desmurs07} (and
with KVN by \cite{Dodson18}). This result solves the discrepancy of the
relative position between the H$_2$O and the SiO maser emission and
offers a consistent picture of the source: The Mira in the center
closely surrounded by SiO maser emission; farther out but related with
the Mira, the H$_2$O masers tracing the base of the outflows also
detected with the thermal SiO emission (this observations) and, far
away, at larger scale, the CO outflows.

\end{document}